# Observation of giant nonreciprocal charge transport from quantum Hall states in a topological insulator

Chunfeng Li[1,2,†], Rui Wang[1,3,†], Shuai Zhang[1,†,*], Yuyuan Qin[1,†], Zhe Ying[1], Boyuan Wei[1], Zheng Dai[1], Fengyi Guo[1], Wei Chen[1], Rong Zhang[2,4], Baigeng Wang[1], Xuefeng Wang[2,*] and Fengqi Song[1,5*]

[1] National Laboratory of Solid State Microstructures, Collaborative Innovation Center of Advanced Microstructures, and School of Physics, Nanjing University, Nanjing 210093, China

[2] Jiangsu Provincial Key Laboratory of Advanced Photonic and Electronic Materials, State Key Laboratory of Spintronics Devices and Technologies, School of Electronic Science and Engineering, Collaborative Innovation Center of Advanced Microstructures, Nanjing University, Nanjing 210093, China

[3] Hefei National Laboratory, Hefei 230088, China

[4] Department of Physics, Xiamen University, Xiamen 361005, China

[5] Institute of Atom Manufacturing, Nanjing University, Suzhou 215163, China

[†] These authors contributed equally to this work.

[*]Corresponding authors. S.Z. (szhang@nju.edu.cn), X.W. (xfwang@nju.edu.cn), F.S. (songfengqi@nju.edu.cn)

**Abstract**

Symmetry breaking in quantum materials is of great importance and can lead to nonreciprocal charge transport. Topological insulators provide a unique platform to study nonreciprocal charge transport due to their surface states, especially quantum Hall states under external magnetic field. Here, we report the observation of nonreciprocal charge transport mediated by quantum Hall states in devices composed of the intrinsic topological insulator Sn-Bi$_{1.1}$Sb$_{0.9}$Te$_2$S, which is attributed to asymmetric scattering between quantum Hall states and Dirac surface states. A giant nonreciprocal coefficient of up to $2.26 \times 10^5$ A$^{-1}$ is found. Our work not only reveals the properties of nonreciprocal charge transport of quantum Hall states in topological insulators, but also paves the way for future electronic devices.

Symmetry is an essential concept in condensed matter physics, and symmetry breaking gives rise to various transport phenomena, such as nonreciprocity. In quantum materials with inversion symmetry breaking, nonreciprocal charge transport phenomena have been observed over the past few years[1,2], such as superconducting diode effects[3-5] and nonlinear Hall effects[6-10]. Nonreciprocal transport manifests in a current direction-dependent resistance, i.e., a difference in the resistance with positive and negative currents, which is crucial in current rectification. When the time-reversal symmetry is broken, a different nonreciprocal response is triggered, which is known as the electrical magnetochiral anisotropy (MCA) effect[11]. To date, numerous nonreciprocal transport phenomena, which are usually probed by the second-harmonic signal under an a.c. current, have been observed[12-33] in inversion-symmetry-breaking materials under external magnetic fields. The various nonreciprocal transport responses in quantum materials have not only shown potential broad applications in electronic components but also deepened the understanding of correlated electronic states[1,2].

In three-dimensional topological insulator (TI) systems[34], nonreciprocal transport has been observed, including bilinear magnetoelectric resistance[35-37] and unidirectional magnetoresistance[25]. Recently, quantum anomalous Hall (QAH)-mediated nonreciprocal transport is observed in magnetic TIs[38,39]. The non-trivial band and ferromagnetism leads to the chiral edge state, and then gives rise to nonreciprocal transport. But the nonreciprocal transport is not large enough, and the magnetism may also contribute to the nonreciprocal transport of distinct origins[25-27]. In an intrinsic nonmagnetic TI, the quantum Hall (QH) effect of topological surface state (TSS) has been realized[40-42]. Due to the similarity with the QAH effect, one may expect that the QH state could provide a good platform exhibiting nonreciprocal transport. However, the QH state has a physical nature distinct from that of the QAH state, as it originates from the quantized Landau levels (LLs) under magnetic fields. This poses an important question, i.e. whether the nonreciprocal transport could be realized in such a system, and what is the possible mechanism if the nonreciprocity indeed takes place.

In this work, we report the observation of giant nonreciprocal transport mediated by QH states in intrinsic TI Sn-Bi$_{1.1}$Sb$_{0.9}$Te$_2$S devices, which is much larger than previous works. And it exhibits a pair of peak and valley with changing gate voltage. Our theoretical calculations suggest that at partial fillings, the asymmetric scattering between the QH edge state and Landau orbits leads to nonreciprocal transport, and the obtained results are in good agreement with the experiments. Our work demonstrates QH-mediated nonreciprocal transport in TIs. It may deepen our understanding of nonreciprocal transport and unexplored phenomena in QH systems at partial fillings.

**Results**

**Observation of QH-state-mediated nonreciprocal transport**

We fabricated devices of Sn-Bi$_{1.1}$Sb$_{0.9}$Te$_2$S (Sn-BSTS)[42,43], a prototypical intrinsic TI with a highly insulating bulk state, by electron beam lithography and electron beam evaporation. Hexagonal boron nitride (h-BN) and graphite were transferred on top of Sn-BSTS by dry transfer. Figure 1a shows the structure of the dual-gated TI device and the configuration of the transport measurement. Graphite/h-BN is used as the top gate

($V_{tg}$), and SiO$_2$/Si is used as the back gate ($V_{bg}$). We define the left and right longitudinal resistances ($R_{xx}$) as illustrated in Fig. 1a. An optical micrograph of a TI device (marked as Device 4) before transfer of the top gate graphite/h-BN is shown in Fig. 1b. The height of Sn-BSTS measured by atomic force microscopy is approximately 39 nm (Supplementary Fig. 1). The temperature ($T$)-dependent $R_{xx}$ of the device in Fig. 1c shows the typical characteristics of an intrinsic TI. As the temperature decreases, the resistance first increases and then decreases, indicating that the TSS is dominant at low temperatures[40]. The temperature-dependent resistance of other devices shows similar behaviour (Supplementary Fig. 2). In the TI device, the QH state can be realized when the Fermi level of the top and bottom surface is tuned into the gap of LLs by top and back gate, respectively. Figure 1d shows the QH state at $T$ = 2 K when $V_{tg}$ is 0.2 V and $V_{bg}$ is 45 V. The Hall resistance ($R_{xy}$) becomes well-quantized when the magnetic field ($B$) is more than 5 T. The quantized Hall resistance is 0.99 $h/e^2$ with a longitudinal resistance of less than 0.006 $h/e^2$, which indicates the high quality of the QH state.

Then, the nonreciprocal resistance $R_{2\omega}$ was measured, which is defined as the second-harmonic resistance $R_{2\omega} = V_{2\omega}/I_{ac}$. The $R_{2\omega}$ values of the left and right sides under the same conditions as in Fig. 1d are shown in Fig. 1e. At a low magnetic field, $R_{2\omega}$ is negligibly small. When the magnetic field increases, the right $R_{2\omega}$ quickly increases and exhibits a peak. When the magnetic field is larger than 5 T, $R_{2\omega}$ decreases to almost zero. In the well-developed QH state, the Fermi level is in the gap of LLs, and chiral edge states occur at the edges of the sample, while the surface is insulating. When the Fermi level is tuned by the gate voltage to slightly deviate from the well-developed QH state, the surface state starts to contribute to transport. Thus, the coexistence of the surface state and QH edge state[38,39] leads to nonreciprocal transport. In addition, the behaviour of the left $R_{2\omega}$ is opposite to that of the right $R_{2\omega}$, which is a signature of chirality of the QH edge states.

From a phenomenological point of view, the current-voltage relationship of the nonreciprocal transport can be written as $V = R_0 I + \gamma R_0 I^2 (\hat{\mathbf{P}} \times \mathbf{B}) \cdot \hat{\mathbf{I}}$, $\hat{\mathbf{P}}$ is a unit vector characterizes the axis of the nonreciprocal effect, $\hat{\mathbf{I}}$ is the current direction, and $\gamma$ is a coefficient to characterize the strength of the nonreciprocal resistance[28,29,38,39]. The second-harmonic voltage $V_{2\omega}$ (the second term on the right-hand side of the equation) is proportional to $I_{ac}^2$. The parabolic dependence behaviour of the current-dependent $V_{2\omega}$ can be clearly seen in Fig. 1f (also in Supplementary Fig. 4). When the a.c. current becomes large, the second-harmonic voltage $V_{2\omega}$ deviates from the parabolic curve. The critical currents here are all larger than 1 μA, so the transport is measured with a current of 1 μA to obtain a better signal-to-noise ratio here.

**Gate controlled nonreciprocal resistance of the top surface**

To gain more insight into the nonreciprocal transport, the dual gate dependent behaviour was studied. In Fig. 2a, b, the mapping of $R_{xx}$ and $R_{xy}$ at $B = -10$ T and $T = 2$ K is shown. The plateaus in Fig. 2a vary from $v = 1$ to $v = 2$ and finally to $v = 3$ with $V_{tg}$ changing from 0 V to 8 V. Here, the bottom surface remains quantized, and the Fermi level of the top surface is tuned to cross the LLs. Correspondingly, the filling factor of the bottom surface is remained at $v_b = 1/2$, and $v_t$ (filling factor of the top surface) is tuned by the top gate[41] from $v_t$ =1/2 to 3/2 to 5/2. Thus, only top surface is modulated here.

The nonreciprocal resistance $R_{2\omega}$ was measured simultaneously, as shown in Fig. 2c. It occurs in the QH plateau transition region obviously, indicated by the red and blue parts of the mapping. To see it more clearly, cuts from the mappings are shown in Fig. 2d-f. In Fig. 2d, e, the green bars indicate the well-developed QH states. From the cut of $R_{2\omega}$ mapping in Fig. 2f, $R_{2\omega}$ is almost zero in the well-developed QH region, and the peak and valley of the gate-dependent $R_{2\omega}$ are located in the QH plateau transition region, as indicated by the arrows. So we can conclude that the QH state itself does not contribute to the nonreciprocal transport, consistent with the magnetic field dependent behaviour in Fig. 1e. At high magnetic field, the QH state is well developed when $R_{xx}$ is minimal, and there are no Dirac surface states. In this case, there is no nonreciprocal behaviour. In contrast, when $R_{xx}$ is maximal, the Fermi level lies in the centre of the LLs, and $R_{2\omega}$ is negligible small. Note that the bottom surface is well quantized, which does not contribute to the nonreciprocal resistance. Here, the transport reflects the nonreciprocal behaviour of the top surface in TI.

In addition, the nonreciprocal resistance in the $v$ =1 to $v$ =2 transition region is larger than that in the $v$ =2 to $v$ =3 transition region. For the high-order plateau transition, the Fermi level is tuned away from the charge neutral point (CNP), and the residual bulk state may contribute to the transport. Due to the bulk inversion symmetry, this state would not contribute to the nonreciprocal transport. Instead, it may attenuate the nonreciprocal resistance[35]. Thus, an insulating bulk state is essential for giant nonreciprocal transport in TIs. In other TI devices, the nonreciprocal behaviour is well reproduced (Supplementary Fig. 6).

**Magnetic field dependent nonreciprocal resistance**

Figure 3 shows the top gate voltage-dependent transport at various magnetic fields at $V_{bg}$ = 44 V. From the mapping of $R_{xx}$ and $R_{xy}$ (Fig. 3a, b), the evolution of QH states can be clearly seen. The dashed lines in the mapping (Fig. 3a, b) is the transition position of the QH state. Due to the back gate is fixed, they indicate the LLs of the top surface. The mapping of $R_{2\omega}$ with top gate voltage and magnetic field is shown in Fig. 3c. It is clear that the nonreciprocal resistance exists only near the LLs. The behaviour of $R_{2\omega}$ is opposite under opposite magnetic fields. As the magnetic field decreases, the valley-to-peak behaviour at a large positive magnetic field (or the peak-to-valley behaviour at a large negative magnetic field) of $R_{2\omega}$ disappears, and there is only a valley at a small positive magnetic field (or a peak at a small negative magnetic field). This is due to the emergence of dissipation at low magnetic fields. From the QH mapping in Fig. 3a, b, $R_{xy}$ deviates from the high-quality plateau and $R_{xx}$ starts to be large in the QH region at low magnetic fields, implying non-negligible dissipation. The additional dissipative channels may smear the valley-to-peak behaviour of the nonreciprocal resistance. This is consistent with the reduced $R_{2\omega}$ in the high-order plateau transition region in Fig. 2f. We also measured magnetic-field-direction dependent $R_{2\omega}$, which satisfies the sinusoidal function fitting (Supplementary Fig. 8). It indicates that only the perpendicular component of the magnetic field is essential in the nonreciprocal transport.

**Giant nonreciprocal transport coefficient**

The nonreciprocal transport coefficient $\gamma$ is of particular interest. To obtain the coefficient, the temperature-dependent $R_{2\omega}$ mapping is measured at $B = -10$ T and $V_{bg} = 44$ V, as shown in Fig. 4a (see Supplementary Fig. 7 for temperature dependent QH mapping). The peak and valley amplitudes of $R_{2\omega}$, indicated by the arrows with the corresponding colours in Fig. 2f, are plotted in Fig. 4b. They all obviously decrease with increasing temperature. $R_{2\omega}$ in the $v=2$ to $v=3$ plateau transition region disappears when the temperature is above 20 K. The high-order ($v > 2$) QH state is also destroyed at high temperatures. The behaviours of the QH state and nonreciprocal transport are highly consistent with each other, i.e. the nonreciprocal resistance can survive to higher temperatures simultaneously with QH states.

We focus on the nonreciprocal resistance in the $v=1$ to $v=2$ plateau transition region at low temperatures, which is the largest in this work. The value of $\gamma$, defined as $\sqrt{2}R_{2\omega}/R_0 I_{ac}$ here, is shown in Fig. 4c. At $T = 2$ K, the coefficient $\gamma$ can be as large as $2.26\times10^5$ A$^{-1}$. This value is approximately two orders of magnitude larger than that in QAH-mediated nonreciprocal transport[38,39], which is usually on the order of approximately $10^3$ A$^{-1}$ (Supplementary Table 1). Here, the coefficient is extracted at high magnetic field. The magnetic field dependent coefficient is shown in Supplementary Fig. 9. The coefficient $|\gamma|$ keeps larger than $1\times10^5$ A$^{-1}$ when $|B| > 6$ T, although there is a slight decrease. At low magnetic field, the nonreciprocal transport is more complex due to the additional contribution.

**Discussions**

A three-dimensional TI can be viewed as two layers of two-dimensional Dirac surface state due to the highly insulating bulk state. Under magnetic fields, the Dirac surface state form LLs with degenerate Landau orbits. The low-energy physics projected onto each LL is essentially the same as that of the QH state in two-dimensional electron gas. Thus, in order to investigate the possible origin of the nonreciprocal transport, we consider a general QH state with randomly distributed impurities. The random impurities broaden the LLs and generate the scattering between the QH edge state and the Landau orbits, as illustrated by Fig. 5a. Interestingly, due to the chiral nature of the state, asymmetric scattering that violates the detailed balance takes place (see Supplementary Note 8 and 9 for details). This leads to dissipation from the edge state to the Landau orbits of the surface state, resulting in nonreciprocal resistance characterized by

$$R_{2\omega} = CI\rho'(E_F), \qquad (1)$$

where $C$ is a constant determined by the QH system itself and the impurities, $\rho(E_F)$ is the density of state (DOS) of the Landau orbits in the broadened LL, and $E_F$ is the Fermi energy of the edge state. The numerical simulation shows that $\rho(E_F)$ is of the Gaussian form under the impurity average (upper half of Fig. 5b). $R_{2\omega}$ is then calculated from Eq. (1), as shown in bottom half of Fig. 5b. It is clear that a pair of peak and valley will occur with changing $E_F$, i.e. gate voltage, which is in good agreement with the experimental results (Fig. 2f). This effective gate modulation differs from the previous nonreciprocal transport in TI[29,35] and QAH system[38,39], which is an important manifestation of the QH-mediated nonreciprocal resistance in TI. The slight decrease of $|\gamma|$ with magnetic field can also be explained by Eq. (1). With the decreasing of magnetic field, the broaden width of LL will increase, and the maximum of $\rho'(E_F)$ will decrease, as well as the nonreciprocal resistance. A detailed numerical simulation of the nonreciprocal transport of the edge states using the Landau-Butticker formula is also presented in Supplementary Note 9. Our theoretical model strongly suggests that the nonreciprocal transport may be general in QH systems.

The contribution from only a single surface is considered above, as the schematic diagram shown in Fig. 5c. Then, we discuss the joint contribution from both the top and bottom surface. The two surfaces are decoupled due to the sufficient thickness of TI, and they can be modulated by the top and back gate, respectively. We extract the maximum value of $R_{2\omega}$ ($V_{tg} \approx 2$ V) under different back gate, i.e., the peak value of each cut in Fig. 2c. As shown in the inset of Fig. 5d, the Fermi level of top surface is fixed in the position where QH edge state and surface state coexist, and $\rho'(E_F)$ of the top surface is maximum. While the Fermi level of bottom surface is varied by back gate, it changes the current distribution on top and bottom surface[37], as well as $\rho'(E_F)$ of the bottom surface. Since $R_{2\omega}$ is proportional to current and $\rho'(E_F)$, we can prove that $R_{2\omega}$ can be largely decreased when the $E_F$ of bottom surface is tuned from the gap between two adjacent LLs to the center of LL (see Supplementary Note 10). So the decoupled two surfaces in TI provide an additional means to modulate the nonreciprocal transport.

Furthermore, the nonreciprocal transport mediated by QH states in TIs is quite different from previous works. Firstly, QH-mediated nonreciprocal transport is different from the bilinear magnetoelectric resistance[35] in TI. The bilinear magnetoelectric resistance originates from the non-equilibrium spin current, while the QH-mediated nonreciprocal transport is not associated with the spin texture, and it does not scale linearly with the magnetic field. Secondly, there are significant differences between QH-mediated and QAH-mediated nonreciprocal transport. QAH-mediated nonreciprocal resistance can occur at zero magnetic field with appropriate gate control[38], while the external magnetic field and LLs are essential in our experiment. Besides, QH breakdown can affect the behavior of the nonreciprocal transport (Supplementary Fig. 5). Importantly, a giant nonreciprocal coefficient is realized through the comprehensive control of magnetic fields and gates, because the derivative of DOS of a broaden LL can be very large. Also, the decoupled two surfaces can increase the nonreciprocal resistance with careful modulation of gates. Besides, the insulating bulk state is important, because conducting bulk state would attenuate the magnitude of nonreciprocal transport. These factors can together affect nonreciprocal coefficient.

For the application in electronics, the magnetic field needs to be lower. The ways to realize the effect at low magnetic field is to optimize the TI material Sn-BSTS[44], or find a new ideal bulk-insulating TI without compensation[45]. Besides, we note that a recent work reports the realization of QAH effect in hundred-nanometer-thick magnetic TI heterostructures[46], which may provide an opportunity to modulate the nonreciprocal transport in QAH system.

In summary, we report the observation of giant QH-mediated nonreciprocal transport in intrinsic TI devices. It originates from the asymmetric scattering between the QH edge state and Landau orbits. The sign of the nonreciprocal resistance is determined by the direction of the magnetic field, edge position, and Fermi level. The magnitude of $\gamma$ is much larger than the previous works in TIs. Our work deepens the understanding of nonreciprocal transport and QH state. It may pave the way for the design of future electronic devices, such as high-performance rectifiers.

## Methods

**Device fabrication.** High-quality Sn-BSTS crystals were grown using the melting method. Thin Sn-BSTS flakes were mechanically exfoliated onto Si substrates with 300 nm thick $SiO_2$. Au electrodes were fabricated by electron beam lithography and electron beam evaporation. The thickness of Sn-BSTS flakes was determined by atomic force microscopy (Cypher S). h-BN and graphite were sequentially transferred on top of the Sn-BSTS flake by dry transfer.

**Transport measurement.** The transport was measured by standard lock-in amplifiers (SRS SR830). An excitation current with a low frequency of 13.333 Hz was applied by a current source (Keithley 6221). For an a.c. current $I = I_0 \sin\omega t$, the voltage is $V = R_0 I_0 \sin\omega t + \gamma R_0 I_0^2 [1 + \sin(2\omega t - \pi/2)]/2$, where the effective value of the applied a.c. current is $I_{ac} = I_0/\sqrt{2}$. The second-harmonic resistance is $R_{2\omega} = \gamma R_0 I_{ac}/\sqrt{2}$, which can be obtained by measuring the a.c. voltage at a frequency of $2\omega$. A Keithley 2400 was used to apply the dual gate voltage. The transport measurements were performed in a Cryomagnetics cryostat with a magnetic field up to 12 T and a temperature down to 1.6 K. The data shown in the main text are raw data without symmetrization or antisymmetrization.

## Data Availability

All data supporting the findings of this study are available within the article and its Supplementary Information. Additional data are available from the corresponding authors upon reasonable request. Source data are provided with this paper.

**Acknowledgements**

This work was supported by the National Key R&D Program of China (No. 2022YFA1402404), the National Natural Science Foundation of China (Grant Nos. 92161201, T2221003, 12374043, T2394473, T2394470, 12322402, 62274085, 12104220, 12025404, and 61822403), and Innovation Program for Quantum Science and Technology (Grant No. 2021ZD0302800)

**Author Contributions**

S.Z., X.W. and F.S. supervised the project. C.L fabricated the devices. C.L and S.Z carried out the transport measurement. R.W. devised the theory. S.Z. R.W., C.L, Y.Q., X.W. and F.S. performed analysis and wrote the manuscript with contributions from all the authors.

## Figures

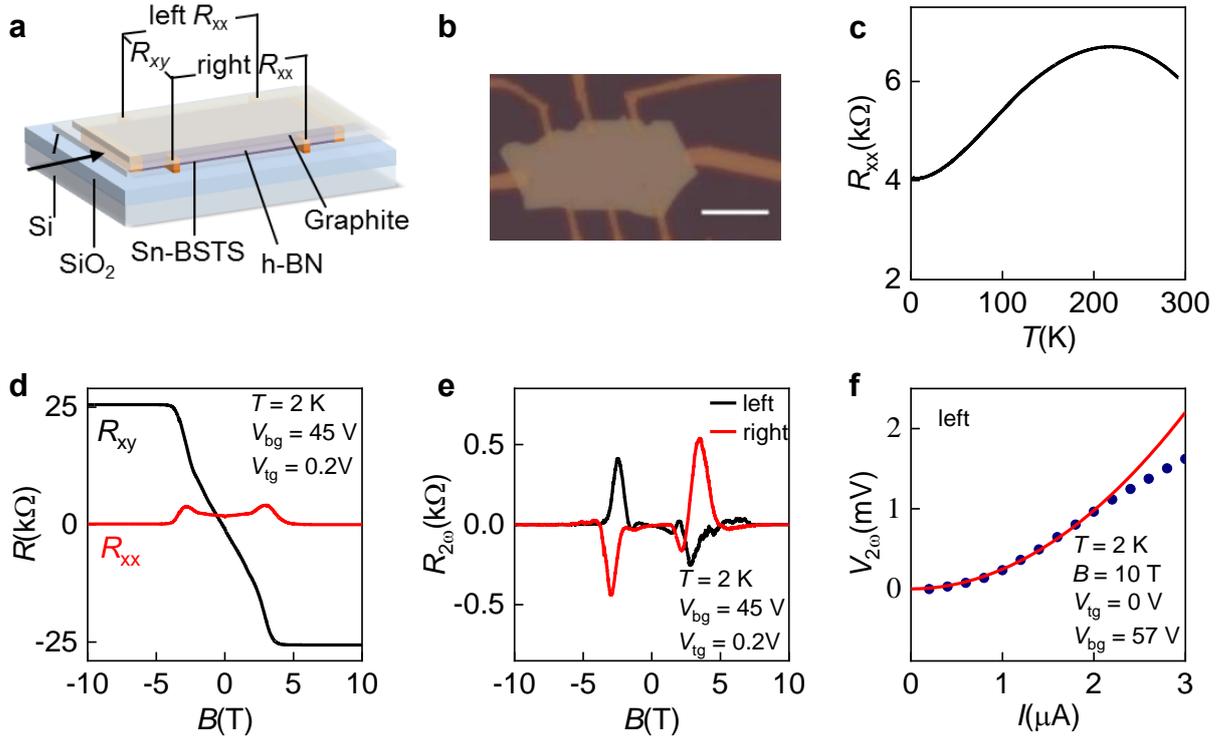

**Fig. 1 | QH-mediated nonreciprocal charge transport in Sn-Bi$_{1.1}$Sb$_{0.9}$Te$_2$S device.**
**a**, Structure of the dual-gated TI device and measurement configuration. The left (right) $R_{xx}$ was measured at the left (right) contacts defined by the injected current direction. Si/SiO$_2$ and graphite/h-BN work as back and top gates, respectively. **b**, Optical micrograph of Device 4 with a thickness of 39 nm before transfer of the h-BN on top of the TI. The white scale bar is 10 μm. **c**, Typical temperature dependent resistance of the TI device. **d**, Magnetic field dependence of $R_{xx}$ and $R_{xy}$ measured at $T$ = 2 K with a top gate voltage of 0.2 V and a back gate voltage of 45 V. A QH state with low dissipation is observed when the magnetic field is larger than 5 T. **e**, Left and right nonreciprocal resistances under the same conditions as in Fig. 1d. $R_{2\omega}$ is opposite under positive and negative magnetic fields, and the left $R_{2\omega}$ is opposite to the right $R_{2\omega}$. This indicates the chirality of the QH edge states. **f**, Current magnitude dependence of the second-harmonic voltage drop $V_{2\omega}$ in the TI device.

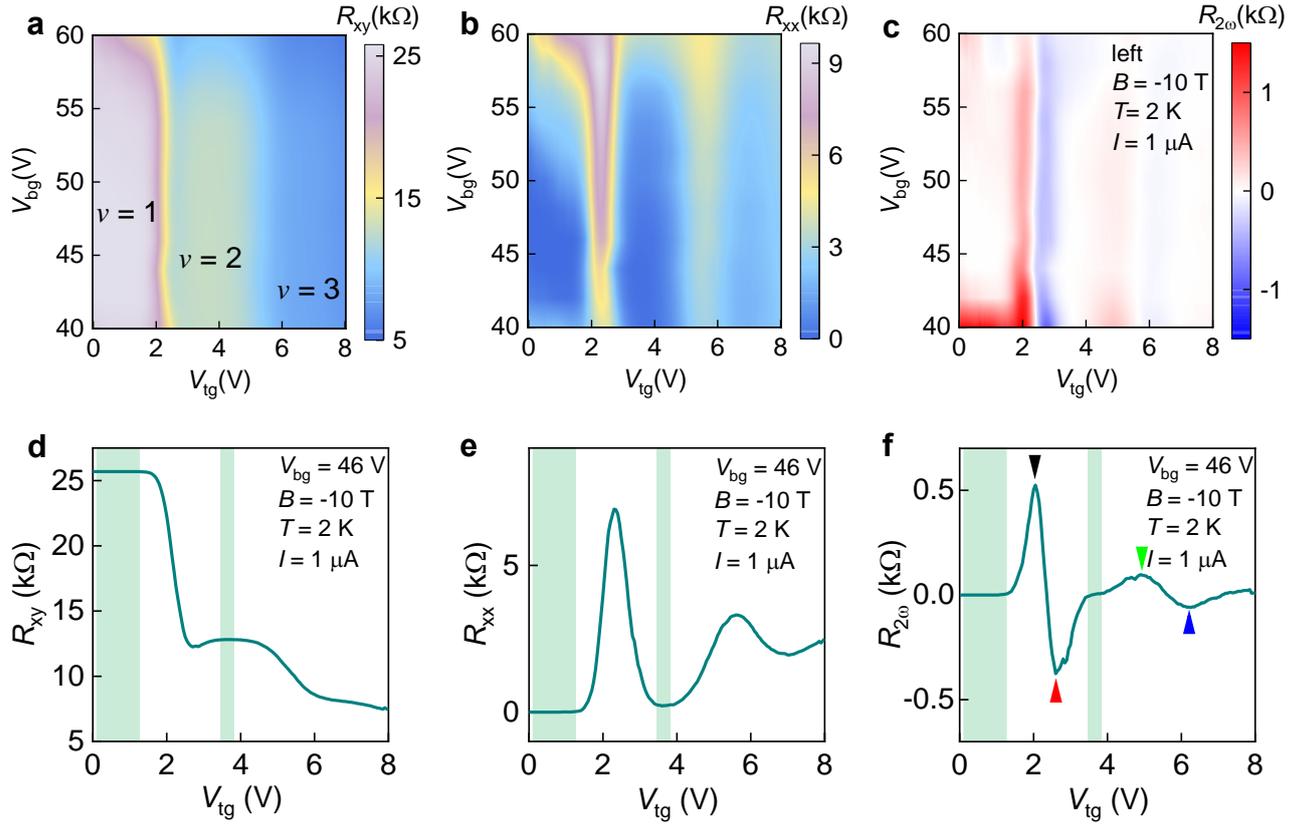

**Fig. 2 | Gate dependent nonreciprocal charge transport in Sn-Bi$_{1.1}$Sb$_{0.9}$Te$_2$S device.** **a, b**, Dual-gated QH mapping at $T = 2$ K and $B = -10$ T with $I_{ac} = 1$ μA. The plateaus vary from v =1 to v =2 and finally to v =3. (**a**) $R_{xy}$ mapping, and (**b**) $R_{xx}$ mapping. **c**, Mapping of the nonreciprocal resistance $R_{2\omega}$ under the same conditions. **d-f**, Top gate voltage dependence of the QH state (**d, e**) and nonreciprocal resistance (**f**) at $V_{bg} = 46$ V, which are cuts from **a-c**. The wide and narrow green bars indicate the well-developed QH states of $v = 1$ and $v = 2$, respectively. The arrows indicate the peak and valley of gate-dependent $R_{2\omega}$, which are located in the QH plateau transition region.

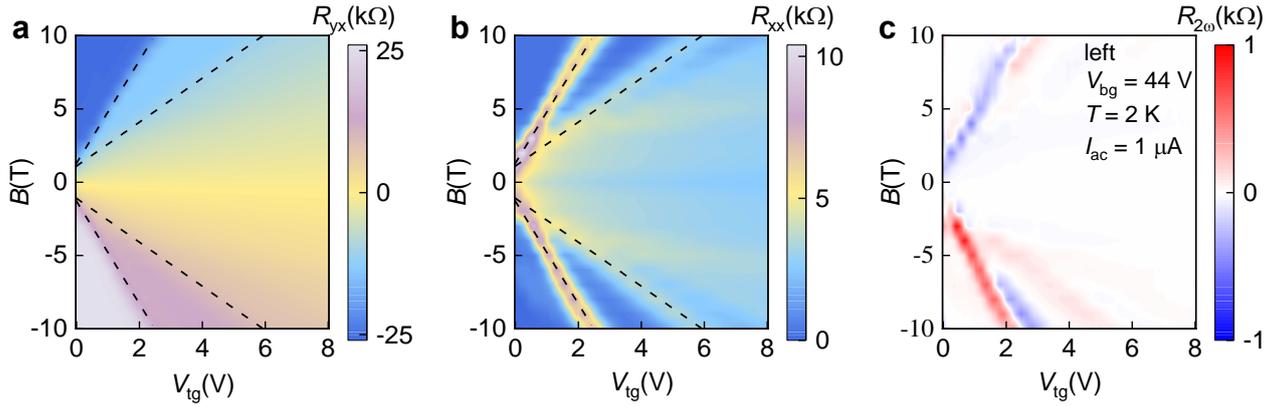

**Fig. 3 | Magnetic field-dependent nonreciprocal resistance in Sn-Bi$_{1.1}$Sb$_{0.9}$Te$_2$S device**. **a, b**, Top gate dependence of (a) $R_{xy}$ and (b) $R_{xx}$ at various magnetic fields at $T$ = 2 K and $V_{bg}$ = 44 V. The dashed lines indicate the Landau levels of the top surface. **c**, Top gate voltage dependence of the nonreciprocal resistance at various magnetic fields under the same conditions. This shows that the nonreciprocal resistance exists near the Landau levels.

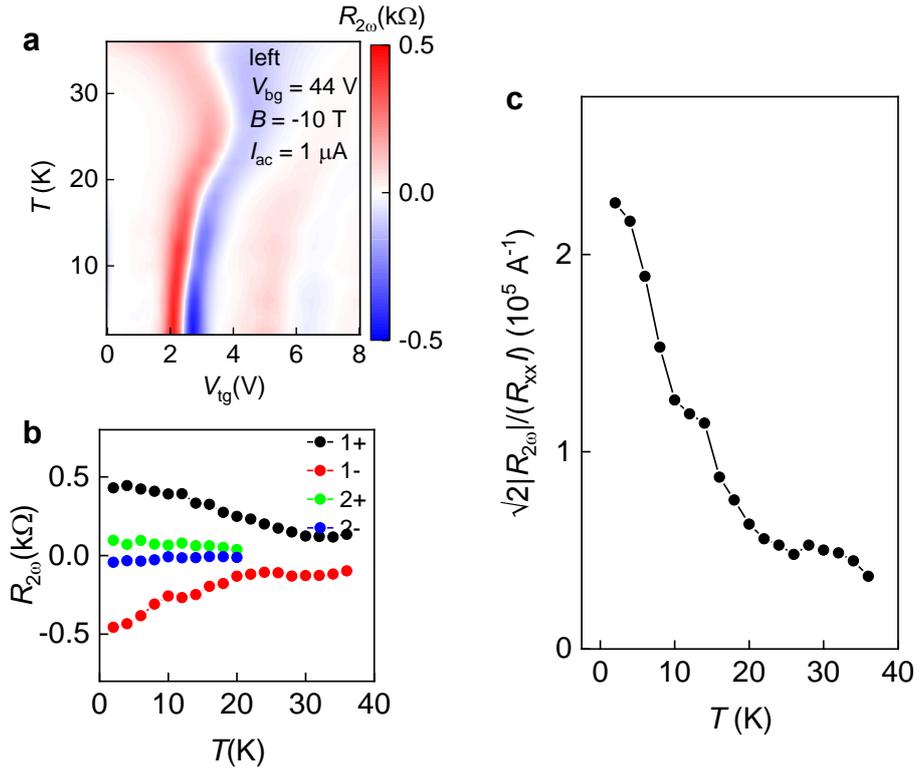

**Fig. 4 | Temperature-dependent nonreciprocal charge transport in Sn-Bi$_{1.1}$Sb$_{0.9}$Te$_2$S device**. **a,** Top gate voltage dependence of the nonreciprocal resistance at various temperatures at $B = -10$ T and $V_{bg} = 44$ V. **b,** Temperature-dependent peak and valley values of the nonreciprocal resistance extracted from **a**. **c**, Magnitude of $\gamma$ at different temperatures. It decreases with increasing temperature and is on the order of $10^5$ A$^{-1}$, which is much larger than that in QAH-mediated nonreciprocal transport.

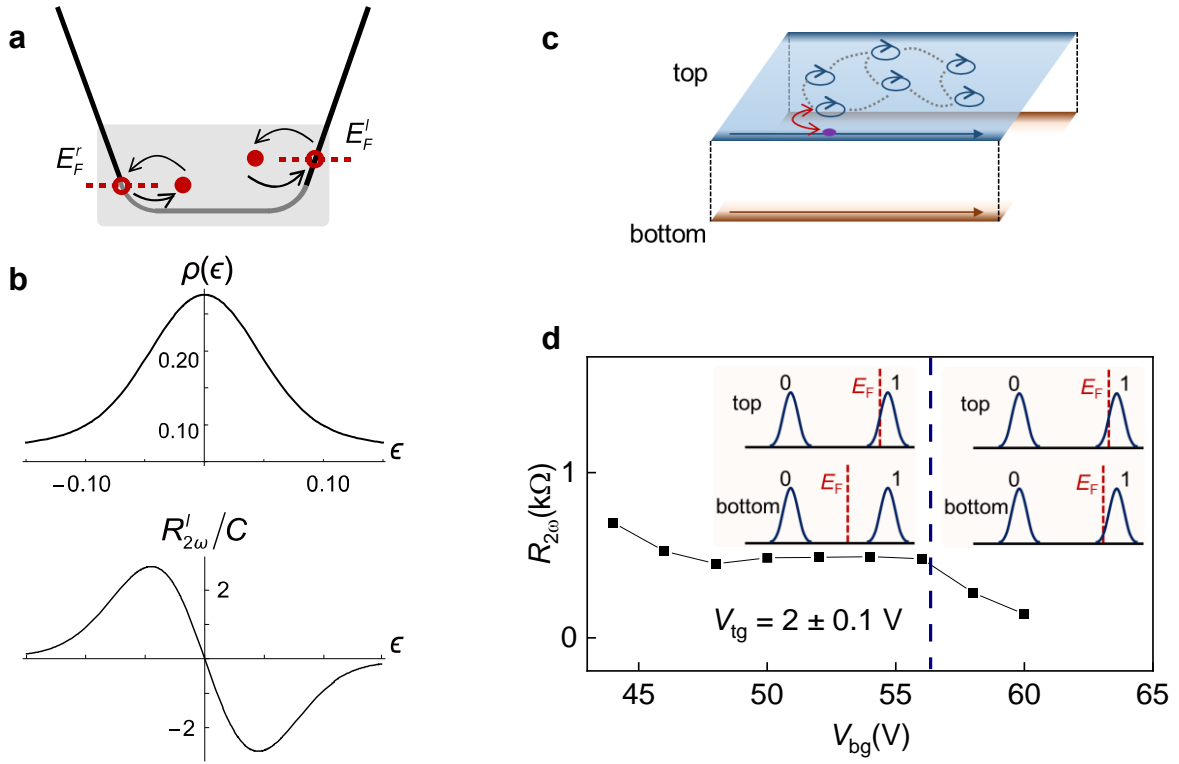

**Figure 5 | Origin of nonreciprocal resistance**. **a**, In presence of boundaries, the edge state would couple to the Landau orbits via impurity scattering. The grey region indicates the broadening of Landau levels by disorder. **b**, The upper half is the density of states corresponding to a single Landau level. The Landau level is broadened due to the impurity scattering. The bottom half is the calculated nonreciprocal resistance, which demonstrates a pair of peak and valley with changing Fermi level, in well consistence with experiments. **c**, Schematic diagram of the nonreciprocal transport from a single surface. The bottom surface is well quantized, while QH edge state and surface state coexist in the top surface, which gives rise to nonreciprocal transport. **d**, Back gate dependent nonreciprocal resistance (extracted from Fig. 2c). The inset shows the position of the Fermi level (red dash line) of the top and bottom surface. The magnitude of nonreciprocal resistance becomes small when the Fermi level of two surfaces on the same level.